\documentclass[sensors,journal,submit,pdftex,moreauthors]{Definitions/mdpi}

\renewcommand{\linenumbers}{}

\firstpage{1}
\pubvolume{1}
\issuenum{1}
\articlenumber{0}
\pubyear{2026}
\copyrightyear{2026}
\datereceived{ }
\daterevised{ } \dateaccepted{ }
\datepublished{ }

\usepackage[normalem]{ulem}
\definecolor{addgreen}{rgb}{0.00,0.45,0.10}
\definecolor{repblue}{rgb}{0.00,0.20,0.75}
\definecolor{cutred}{rgb}{0.70,0.10,0.10}

\Title{An Assessment of the Triangulation Capabilities of a Multi-Site All-Sky Infrared Camera Array}

\Author{Abraham Loeb $^{1,2}$, Alex Delacroix $^{1,2}$, Ezra Kelderman $^{1,2}$, Eric Masson $^{1,2}$, Regina Sarmiento $^{1,2}$}

\AuthorNames{Abraham Loeb, Alex Delacroix, Ezra Kelderman, Eric Masson, Regina Sarmiento}

\address{$^{1}$ \quad Harvard-Smithsonian Center for Astrophysics, 60 Garden Street, Cambridge, MA 02138, USA\\
$^{2}$ \quad Galileo Project, 60 Garden Street, Cambridge, MA 02138, USA}

\corres{Correspondence: richardcloete@cfa.harvard.edu}

\abstract{The Galileo Project operates three autonomous multi-camera observing sites near Las Vegas, Nevada, separated by baselines of $2$ and $10\,$km, with the goal of characterizing the populations of natural and human-made objects that cross the sky and thereby isolating any event that fits none of them. Each site carries an array of eight long-wave infrared (LWIR) cameras arranged in a fixed dome-like structure, an IR-Dalek, providing $360^{\circ}$ azimuthal coverage. We report the commissioning of the three arrays and present the triangulation pipeline built on them. Because our targets are not assumed to follow any particular trajectory, we depart from the parametric fits standard in meteor work and estimate the instantaneous three-dimensional position at each time step as the point that minimizes the weighted sum of squared distances to the simultaneous lines of sight, weighting each camera by its own pointing uncertainty and by its range to the object. We describe the intrinsic and extrinsic calibration procedure, which uses Automatic Dependent Surveillance--Broadcast (ADS-B) aircraft as reference sources in the absence of stars or fixed landmarks in the LWIR field of view, and the multi-sensor track association scheme that decides which detections belong to the same object. Using one week of data (2026 May 24--30; $1{,}650$ camera-hours, $2.1\times10^{8}$ detections and $5.3\times10^{6}$ tracks), we validate the recovered positions, speeds and accelerations against ADS-B records, recovering distances to within $5$\% for $99$\% of the aircraft solved from three sites and for $83$\% of those solved from a single site pair and speeds to within $5$\% for $78$\% and $63$\% of them respectively. We characterize the identification completeness of the array, which is set by apparent size: the $50$\% identification threshold falls at $1.8$ pixels of apparent wingspan.
}

\keyword{unidentified aerial phenomena; triangulation; multi-site sensor networks; long-wave infrared imaging; camera calibration; ADS-B; track association; trajectory reconstruction}

\begin{document}

\section{Introduction}

In the effort to obtain science-quality data for the analysis of unidentified aerial phenomena (UAP), the Galileo Project has designed and deployed five observing sites to monitor the skies \citep{doi:10.1142/S2251171723400032}. The project aims to characterize the overall distribution of human-made and natural phenomena, so as to distinguish any anomalous event that does not fit any of these known populations \citep{doi:10.1142/S2251171723400068}.

In this work we focus on the three sites located in Las Vegas, Nevada (United States). Each of these sites has an array of eight long-wave infrared (LWIR) wide-field cameras that continuously observe the sky. Four visible wide-field cameras per site are installed but not yet commissioned, and are not used here. Each site operates autonomously, with onboard edge computing, local data storage, GPS-disciplined timing, and automated thermal management. Separated by baselines of $2\,$km and $10\,$km, this configuration enables the combination of detections from the different sites to reliably determine 3D positions, velocities and accelerations of an object that is simultaneously detected by two different sites. Characterizing the triangulated 3D trajectories allows us to further separate the dynamical properties of known objects (e.g. solar system bodies, aircraft, clouds) in the search for outliers.

The triangulation of 3D positions from directional data has been extensively explored for meteor trajectory reconstruction. Pioneer works like \cite{1987BAICz..38..222C} proposed that a meteor's trajectory can be estimated as the intersection of the planes formed by the sites and the lines of sight of the source observed at different times, from each site. To avoid reconciling different planes when more than two sites observe the event, \cite{1990BAICz..41..391B} proposed an approach that naturally combines multi-site observations. This consists of finding the linear trajectory that minimizes the squared distances between the different lines of sight and the trajectory. In more recent works, instrument uncertainties and distances to the observation sites have been introduced to weight the directional contributions (\citealt{2019A&A...627A..78J}, \citealt{2021MNRAS.506.5046V}, \citealt{2022A&A...663A.154M}). A common assumption across all of these works is that the observed objects describe linear trajectories. This is a valid approximation for bolides and a reasonable choice when simultaneous data points are not available (\cite{2012M&PS...47.1405G}).

Since our goal is to track arbitrary trajectories, we depart from these approaches by computing the object's instantaneous position at each time step, rather than fitting a parametric trajectory. We leverage our multi-site synchronized setup to approximate the object's instantaneous position by finding the 3D position that minimizes the sum of the weighted squared distances from all visual rays, at a given time. This weighted approach takes into account the direction uncertainties intrinsic to our observing system and the distance from which each site observes the object. This choice yields noisier per-point estimates, which we address in Sec.~\ref{sec:proper_triangulation} by fitting a smooth curve to the reconstructed track.

This work continues the triangulation analysis projected in \cite{doi:10.1142/S2251171723400020}, and follows our first commissioning experience with our development site in Massachusetts \citep{s25030783}.

Building on the recent availability of LWIR data from these sites, we present the commissioning results of the camera arrays installed at each site (Sec.~\ref{sec:commissioning}), and illustrate our triangulation approach using the LWIR-instruments across the network. For this, we define our triangulation method (Sec.~\ref{sec:triangulation_math}) and the multi-sensor track association approach (Sec.~\ref{sec:track_association}). We test our triangulation calculations against ADS-B aircraft records (Sec.~\ref{sec:proper_triangulation}), and close by discussing the selection effects the array imposes and the limits they place on any search for anomalous objects (Sec.~\ref{sec:discussion}).

\section{Materials and Methods}

The Galileo Project has installed 3 observing sites in Nevada, to which we refer as Sphere (Lat:\,36.121, Lon:\,-115.161), Hideout (Lat:\,36.072, Lon:\,-115.066) and Wildhorse (Lat:\,36.057, Lon:\,-115.079). Sphere lies $10.1\,$km from each of the other two, which are only $2.0\,$km apart, so the array is strongly asymmetric: the Hideout--Wildhorse pair constrains range far more weakly than either does against Sphere. Each of these stations is equipped with long wave infrared (LWIR) wide angle cameras that continuously record the skies of each site $10^{\circ}$ above the horizon. Each site operates autonomously, with onboard edge computing, GPS-disciplined timing, and automated thermal management. In one of the sites, Hideout, an ADS-B receiver (RTL-SDR dongle) tracks aircraft positions.

In this section, we give an overview of the instruments, the data they produce, their calibration, and the software pipeline implemented to obtain triangulated positions.

\subsection{Instruments}

\subsubsection{The IR-Dalek}

The LWIR cameras are arranged in dedicated fixed structures called Daleks to cover different portions of the sky. The Daleks in Las Vegas sites replicate the first-generation Dalek, which was commissioned at a Massachusetts observatory in February~2025 and is described in detail in \cite{s25030783}. We briefly describe the Dalek set up.

\begin{figure}[t]
\centerline{\includegraphics[width=0.9\textwidth]{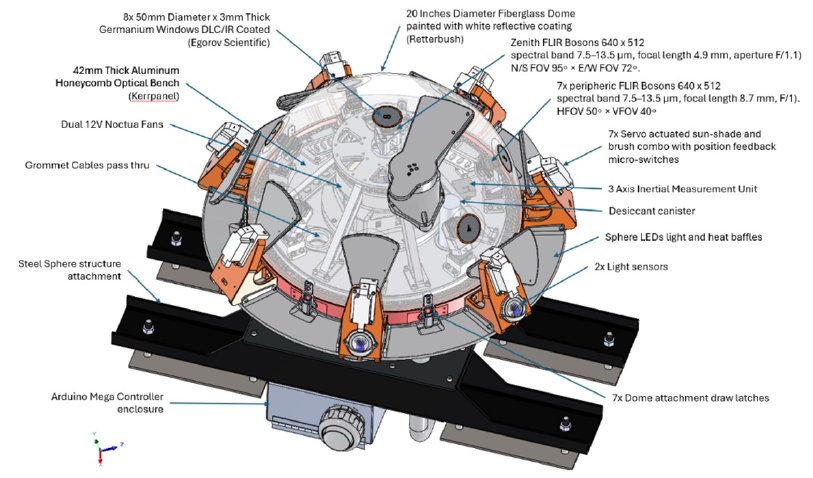}}
\caption{IR-Dalek: the eight-camera LWIR array at the Sphere site. Seven cameras point outwards around a circle at $\sim30^{\circ}$ elevation and an eighth, wider-field camera points at the zenith.}
\label{fig:dalek}
\end{figure}

The IR-Dalek has eight FLIR LWIR Boson $640\times512$ sensors (spectral band $7.5$ - $13.5\,\mu$m). Seven of these have a $50^{\circ}$ HFoV (focal length $8.7\,$mm, $f/1$) and are distributed on a circle pointing outwards with a $30^{\circ}$ elevation. An eighth IR Boson camera with a {$4.9\,$mm lens} ($f/1.1$) located within the circle is oriented towards the zenith (Fig.~\ref{fig:dalek}). This arrangement results in a $360^{\circ}$ azimuthal and $160^{\circ}$ elevation coverage. For all instruments, the camera array is oriented clockwise from North, with Camera 1 pointing North.

The seven outward-looking cameras and the zenith camera are mounted on a custom rigid aluminum honeycomb optical bench via FP90 Thorlabs flip platforms, allowing accurate $30^{\circ}$ elevation adjustment. A custom 20-inch ($508\,$mm) diameter fiberglass dome protects the cameras from wind, rain, dust, and wildlife.

The Boson cameras view the sky through germanium windows bonded to the dome with UV-grade silicone adhesive, carrying a diamond-like carbon (DLC) anti-reflective coating on the outer surface and an AR coating on the inner surface. Additionally, each camera is equipped with a servo-actuated shade that opens and closes on a per-camera schedule, tracking the sun across the sky, protecting the microbolometer sensors from sustained direct solar exposure.

The shades are each driven by a ZOSKAY DS3218 $270^{\circ}$ full-metal-gear digital servo with closed-state confirmation from a D2HW-C202M normally-closed micro-switch. Shade state is evaluated continuously from solar ephemeris data computed onboard via the \texttt{SolarCalculator} library, synchronized to UTC via NTP, running on an Arduino Mega with Ethernet shield.

While we report some blocked shades and the Sun is seen in some recordings, we do not find any permanent damage in the instruments.

\subsubsection{The Garmin GPS}

\begin{figure}[t]
\centerline{\includegraphics[width=0.9\textwidth]{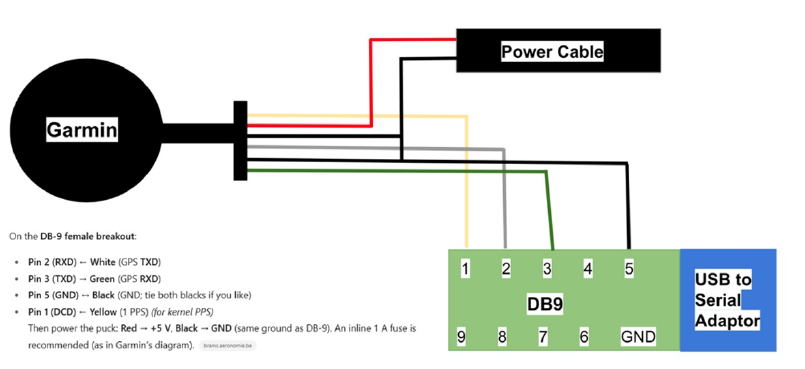}}
\caption{Garmin GPS custom cable and DB9 connector wiring.}
\label{fig:gps_cable}
\end{figure}

Each site has a Garmin GPS receiver, which is connected to the main edge computer at each site via a custom DB9 serial interface and USB-to-serial adapter (Fig.~\ref{fig:gps_cable}). This provides a pulse per second (PPS) GPS-disciplined UTC reference on the main computer. Then, the instrument edge computers are synchronized through the Network Time Protocol (NTP).

The resulting timestamp synchronization between each camera instrument and the GPS-disciplined edge computer has been measured at $0.25$--$0.30\,$ms. Since all instruments synchronize to the same edge computer, the inter-instrument timestamp offset across the full network remains below $\sim0.6\,$ms. This is well within the timing budget imposed by the 60~fps recording rate of the Dalek cameras (frame interval $\sim16.7\,$ms).

\subsubsection{The ADS-B receiver}

For control and calibration purposes, we use ADS-B data. The Hideout ADS-B subsystem consists of a FlightAware USB software-defined radio (SDR) dongle (1090~MHz, Mode-S) connected to a dedicated Raspberry Pi edge computer with a local 1090~MHz antenna and local data storage. The data is stored using the \texttt{readsb} open software \citep{wiedehopfreadsb}. The centralized single-receiver architecture reduces reliance on third-party flight-tracking services and preserves raw packet-level timing.

\begin{figure}[t]
\centerline{\includegraphics[width=0.9\textwidth]{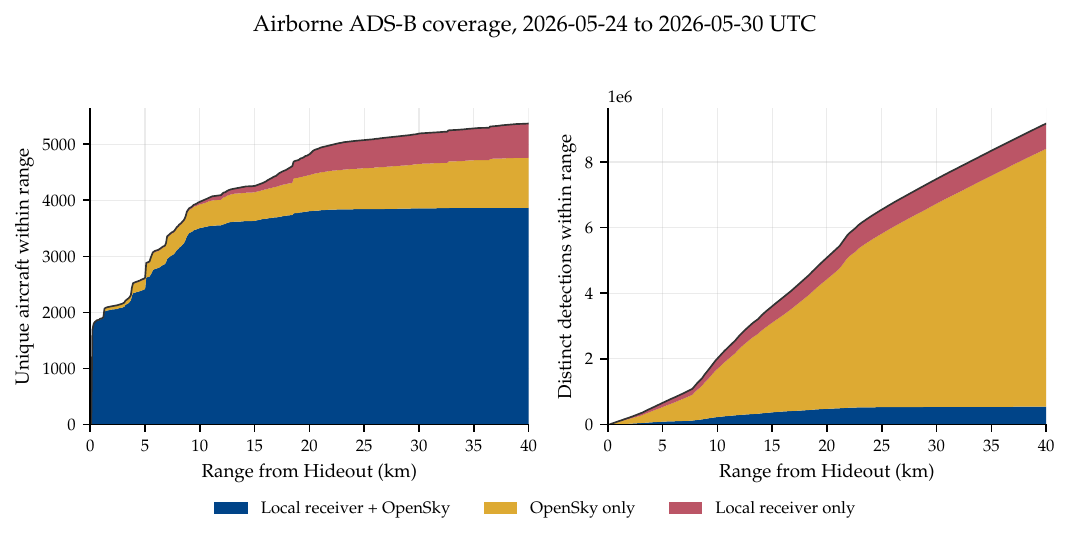}}
\caption{Coverage of the Hideout ADS-B receiver compared with the \texttt{OpenSky} archive over the same period. The local receiver is limited in range by the horizon and by obstructions around the antenna, but recovers aircraft that \texttt{OpenSky} does not.}
\label{fig:adsb_num_coverage}
\end{figure}

Due to physical objects blocking the receiver's FoV and limited range of our receiver, we complement our locally acquired ADS-B data with historical data pulled from \texttt{OpenSky} \citep{schafer2014bringing}. Over the week analyzed here the Hideout ADS-B receiver recovers $81$\% of the aircraft that \texttt{OpenSky} reports within $40\,$km ($4{,}475$ of $5{,}368$ distinct airframes), however it does so with a more limited range (see Fig.~\ref{fig:adsb_num_coverage}). Its share of the individual position reports falls monotonically with range, from $33$\% inside $5\,$km to $9$\% at $20$--$25\,$km and under $3$\% beyond $25\,$km, so \texttt{OpenSky} supplies almost all of the coverage at the far end of the array's useful volume.

The Hideout local receiver also contributes other aircraft data not present in the \texttt{OpenSky} dataset ($610$ airframes over this week). This corresponds to aircraft that do not broadcast their own position, but nearby ground stations identify them and broadcast an additional signal.

\subsection{Data}\label{sec:data}

To illustrate the triangulation pipeline set up for the LWIR-Daleks, we use one week of data: from May 24 through May 30 of 2026 (UTC). During this week, the sites have recorded 1,650 camera-hours, of which $1{,}611$ were processed through the detection pipeline. Each camera has a simultaneous 11-hour-long uptime window per day (downtime through 1 a.m. to 2 p.m. UTC). We exclude from the analysis the recordings corresponding to cameras 3 and 6 at Hideout, and camera 8 in Wildhorse, where we report issues with the shades blocking the field of view over these dates. This leaves 21 of the 24 cameras in the analysis.

The data consists of 15 minutes-long RGB video recordings at 60 frames per second with $640\times512$\,px$^2$ produced using the \texttt{Gstreamer} open media framework \citep{gstreamer}. The recordings are synchronized across the cameras and sites to cover the same time windows.

This raw data has been further processed with our YOLO11-tracking pipeline to identify moving objects in the FoV \citep{doi:10.1142/S2251171723400081}. This output determines the per frame pixel positions of an object within the camera's FoV ($c_x$ , $c_y$), a bounding box size, and a confidence level for the detection ranging from 0 to 1. Additionally, it groups the detections over time identifying distinct tracks. This way, an object observed by one camera is associated to one \footnote{An object can be associated with more than a track when the visibility of the object is discontinuous.} time series of the 2D projected position in the camera's detector.

For the analysis, we filter tracks to keep only those that have at least 30 detections with confidence level greater or equal to 0.5. This is equivalent to half a second of confident detections if they had been seen in consecutive frames. The detector returned $213{,}621{,}108$ detections in total, of which $117{,}378{,}644$ ($55$\%) carry a confidence above $0.5$, grouped into $5{,}313{,}389$ distinct tracks. The length-and-confidence filter above is applied downstream, when the tracks are passed to the identification stage.

As the triangulation sites are located between two and ten kilometers from each other, most of the observed objects are expected to appear different sizes when seen from the different sites. For this reason, we limit the analysis to their positions in the camera's FoV, instead of including any other visual information (e.g. object bounding box size) to aid the object matching.

The combined ADS-B data from \texttt{Opensky} and our local receiver was further processed to derive positions in ECEF, speeds and accelerations. Speeds are taken from the transponder's own reported quantities, $v=\sqrt{v_{\rm gs}^{2}+v_{z}^{2}}$, combining ground speed and vertical rate, rather than by differencing reported positions. Accelerations are obtained by differentiating that reported velocity vector once, rather than by differencing position twice; this lowers the noise floor of the reference from tens of m$\,$s$^{-2}$ to $\sim\!1\,$m$\,$s$^{-2}$. We note that ground speed, track and vertical rate are not refreshed with every position report: reports arrive about once per second, and roughly half of them repeat the previous velocity unchanged, so a large fraction ($83$\%) of the differenced acceleration samples are exactly zero. The fields are also quantized --- $0.1\,$kt in ground speed and $64\,$ft$\,$min$^{-1}$ in vertical rate --- which sets the smallest change the reference can express, but it is the repetition rather than the step size that empties the differenced sample. Where we score acceleration against this reference we therefore restrict to the samples at which the transponder actually reported a change in velocity, which are the only ones that carry information.

\subsection{Calibration}

The calibration of the instruments is essential to accurately establish directions in which an object is observed and at what time, to effectively crossmatch the directions of different simultaneous observations.

While the time sync is achieved synchronizing the instrument's clock to a reliable GPS clock, the cameras require two distinct steps: intrinsic and extrinsic calibrations. The intrinsic calibration models the camera's lens distortions, and the extrinsic calibration positions the camera in an Earth reference system. 

\subsubsection{Dalek cameras calibration}\label{sec:dalek_calibration}

The intrinsic calibration consists of finding the focal length, optical image center and distortion coefficients that model the camera lens. These parameters are approximated by Standard Lens and Fisheye models for the $50^{\circ}$ HFoV and wide-field zenith sensors, respectively. To calculate these parameters, we first image a heated checker board in different angles and distances across the field of view of each camera (between 20-80 images). Then, the images are processed with the \texttt{Open cv2} library \citep{opencv_library} to identify the checker board and calculate the distortions.

For the outward-looking cameras  the angle subtended by a pixel is not constant across the frame: at the mid-field radius of a typical detection it is some $30$\% larger than on the optical axis. We find an intrinsic median re-projection error \citep{doi:10.1142/S2251171723400020} per camera  between $0.5$ and $1.7$ pixels (median $1.1\,$px over the 24 cameras) for the IR cameras.

Once the camera lens is modelled, we can convert pixel measurements ($c_x$ , $c_y$) to unit line-of-sight vectors in the camera frame $\mathbf{d}\propto \mathbf{K}^{-1}(u_x, u_y, 1)^{\mathsf T}$, where $(u_x,u_y)$ are the undistorted pixel coordinates and $\mathbf{K}$ is the intrinsic matrix.

\begin{figure}[t]
\centerline{\includegraphics[width=0.9\textwidth]{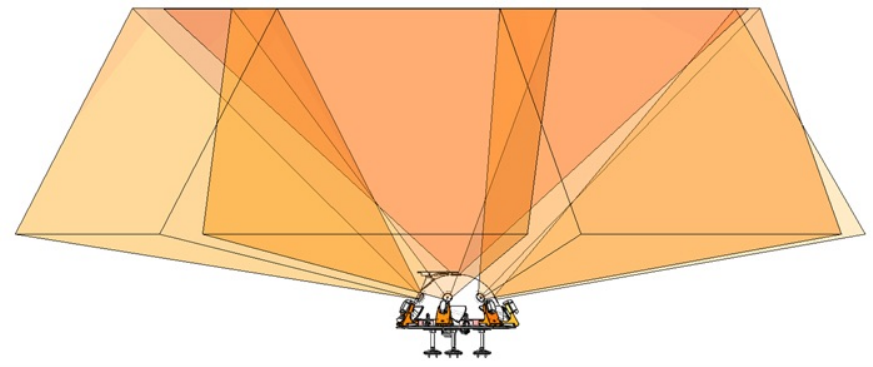}}
\caption{Dalek FoV: the union of the eight cameras' fields of view for a single Dalek, and the overlap between neighboring cameras.}
\label{fig:dalek_fov}
\end{figure}

The extrinsic calibration aims to find the position and orientation of the camera in space and in reference to an Earth reference frame. On the one hand, we need to know where the instrument is and, on the other, where are the cameras pointing at.

To determine the cameras' positions, we use the Dalek's GPS position, which has an expected error of $1-3$\,m. For the spatial orientation, we require a minimum of six observable sources with known positions to determine a rotation matrix that converts between the camera frame to a world frame. Due to the lack of fixed landmarks in the field of view and stars not being observable with the IR-cameras, we rely on aircraft positions tracked by the ADS-B. Finding the rotation matrix $\mathbf{R}$ is a known problem \citep{wahba1965least}, for which we follow the solution procedure described in \cite{doi:10.1142/S2251171723400020} (Sec.\,4.2.1).

Due to unsystematic uncertainty in the ADS-B timestamps (e.g. uncompensated latency of $-0.2$ to $+0.4\,$s, constant along a given aircraft's track but varying between aircraft, \citealt{george2020studyuncompensatedlatencyadsb}), we aimed to maximize the number of identified aircraft tracks to smooth the dependence with tracks timestamp uncertainty. To automate the track identification, we perform the calibration in an iterative fashion.

Initially, we set the rotation matrices to the design camera orientations (Fig.~\ref{fig:dalek_fov}). We then match the observed tracks to the known aircraft ground truth tracks. As the initial estimate might lead to wrong detections and other objects could be confused with aircraft, we proceeded to attempt to triangulate the aircraft as a further verification that the object in sight has been identified correctly. When the aircraft tracks could be recovered with a position error smaller than $1\,km$, we store the observation-ground truth matches and used these to compute the new camera's rotation matrix.

After refining the rotation matrix, some of the cameras may still exhibit large re-projection errors of the selected tracks. The cameras that have a re-projection error lower than five pixels become the trusted cameras that define the structural orientation of the Dalek. When a camera has a median re-projection error that exceeds 5px, we assign it its corresponding structural rotation matrix. At this point, we restart the process by matching the observed tracks with the ADS-B aircraft tracks. We repeat until the cameras show no re-projection error improvement. To minimize the direction uncertainty impact in our calibration, we restrict to aircraft at over $5\,km$ from the instrument. 

Finally, we store the rotation matrix for each camera and the median re-projection error of all their observed identified tracks to estimate the camera pointing uncertainty.

The uncertainty in the triangulated positions lies essentially in the time latency uncertainty and the direction uncertainty of the visual ray and the distance at which the object is from the camera. This includes the intrinsic, extrinsic, and timestamp camera uncertainties, the YOLO consistency at finding the center of the object (if resolved or the image spatial discretization otherwise). Overall, we estimate this error as the difference between where we expect to see the ADS-B position projected on to the camera plane and the detected position. This is the re-projection error. Expressed as an angle, the median extrinsic re-projection residual, converted to a Gaussian $\sigma$ under the Rayleigh assumption of Sec.~\ref{sec:triangulation_math} and divided by the camera's own focal length, corresponds to a pointing uncertainty of $\sim4.7$\,mrad for a typical outward-looking camera, or $\sim56\,$m of transverse position error at $12\,$km.

The extrinsic rotations were fit from 2,645 identified aircraft tracks recorded over the two days immediately before the analysis data (2026 May 22 and 23), a median of 82 tracks per camera. Of the 24 cameras, 21 were solved and the three whose shades stalled closed were excluded; 11 of the 21 reach a median re-projection error below 5\,px on their calibration sample. The final solve replaced the rotation of 16 cameras and kept the previous iteration's rotation for the remaining five, where the new solution did not improve the re-projection error. Further details on the extrinsic calibration results can be found in Appendix~\ref{app:extrinsic-calobration}.

\subsection{Triangulation}

In the context of 2D imaging, the calculation of distance is possible if we can accurately estimate the direction towards the same object as seen in two images taken simultaneously from known (and different) positions. In other words, we need to know the deformation, location and orientation of two cameras, as well as identifying that the object seen from both cameras is in fact the same object. While finding object directions has already been covered in Sec.\,\ref{sec:dalek_calibration}, in this section we focus on the triangulation calculation itself. First, we define how the positions are calculated when an object is tracked by different cameras. Second, we describe the approach taken to identify a unique object and match its tracks observed by the different cameras.

\subsubsection{From object detection to 3D positions, velocities and accelerations}\label{sec:triangulation_math}

Triangulating positions of objects is a geometric problem. We use the Earth-Centered, Earth-Fixed (ECEF) reference frame, in which we locate the sites $\mathbf{S}_1$, $\mathbf{S}_2$ and $\mathbf{S}_3$ (i.e. Sphere, Hideout and Wildhorse). The previous calibration efforts allow us to convert a position of a detection in an image $(x,\,y)$ taken from $\mathbf{S}_i$ to a unitary vector in space $\mathbf{v}_{i}^{ECEF}$ that extends from site $\mathbf{S}_i$ to the detected object:

\begin{equation}
\mathbf{v}_{i}^{ECEF}\;=\;\frac{\mathbf{R}_i^{\mathsf T}\,\mathbf{K}_i^{-1}\,\tilde{\mathbf{u}}_i}
    {\left\lVert \mathbf{R}_i^{\mathsf T}\,\mathbf{K}_i^{-1}\,\tilde{\mathbf{u}}_i \right\rVert}
\end{equation}

\noindent where $\tilde{\mathbf{u}}_i=(u_x,u_y,1)^{\mathsf T}$ are the homogeneous \emph{undistorted} pixel coordinates of the detection, obtained by inverting the intrinsic lens model of Sec.~\ref{sec:dalek_calibration}; $\mathbf{K}_i$ is the intrinsic matrix, whose inverse carries pixels into normalized camera coordinates; and $\mathbf{R}_i$ is the rotation matrix taking world coordinates into the camera frame, so its transpose returns the ray to ECEF. The result is dimensionless by construction and no separate unit conversion is required. We can then express the position of the object from the ECEF origin as

\begin{equation}
    \mathbf{r}_{i}(t) = \mathbf{S}_i + t\,{\mathbf{v}}_{i}^{\text{ECEF}}, \qquad t > 0.
\end{equation}

Given the pixel positions of the same object observed at the same time from different sites, we can project the vectors that point towards the object. Ideally, these would intersect at the position of the object. However, due to uncertainties in the measured directions to the object from the cameras, we estimate the 3D position $\mathbf{X}$ as the point that minimizes the sum of the weighted squared distances from all visual rays as a first order correction.

To calculate the distance from the camera ray $\mathbf{v}_i$, we use the matrix that projects a 3D point to its perpendicular plane

\begin{equation}
    \mathbf{P}_i = \mathbf{I} - \mathbf{v}_{\color{addgreen}i}\,\mathbf{v}_{\color{addgreen}i}^{\mathsf T}
\end{equation}

Then, we minimize the squared sum of such distances across all sites.

\begin{equation}
    J(X) = \sum_i w_i\, \lVert P_i (X - S_i)\rVert^2
\end{equation}

\noindent where the weights $w_i = f_i^2\,(\sigma_i R_i)^{-2}$ are estimated from the camera's focal length $f_i$ for unit conversion to angle, the $\sigma_i$ inferred from each camera's median extrinsic re-projection residual on the assumption that the residuals are Rayleigh distributed, and the distance from the camera to the object $R_i$. Setting $\nabla J = 0$, we obtain the normal equations exactly, since $\mathbf{P}_i$ is a symmetric idempotent projector and $J$ is quadratic in $\mathbf{X}$.

\begin{equation*}
    X = M^{-1} b, \quad
    M = \sum_i w_i P_i, \quad
    b = \sum_i w_i P_i S_i
\end{equation*}

Because $R_i$ depends on the unknown position, the weights are applied in two passes: a first solve with $w_i=1$ supplies the ranges, and a second solve uses $w_i = f_i^2(\sigma_i R_i)^{-2}$. The matrix $\mathbf{M}$ is singular when all rays are parallel, which is the degenerate single-site case and why SSBs positions are poorly constrained in this system. Additionally, solutions are rejected when the mean re-projection error of the recovered point exceeds $10\,$px.

Furthermore, we can estimate the expected error of a given triangulated position $\mathbf{X}$ from the same normal equations. Propagating a perturbation of the ray directions through $\nabla J = 0$ gives $\delta\mathbf{X} = \mathbf{M}^{-1}\sum_i w_i R_i\,
\delta\mathbf{v}_i$, so

\begin{equation}
    \mathrm{Cov}(\mathbf{X}) \;=\; \mathbf{M}^{-1}
    \left( \sum_i w_i^2\,\sigma_i^2\,R_i^2\, \mathbf{J}_i \mathbf{J}_i^{\mathsf T} \right)
    \mathbf{M}^{-1} \;=\; \mathbf{M}^{-1},
\end{equation}

\noindent where $\mathbf{J}_i$ is the Jacobian that maps a pixel perturbation to a ray-direction shift in 3D space. The collapse to $\mathbf{M}^{-1}$ follows because $\mathbf{J}_i \mathbf{J}_i^{\mathsf T} = \mathbf{P}_i/f_i^2$, so the bracket reduces to $\sum_i w_i \mathbf{P}_i = \mathbf{M}$. This holds \emph{only} under the inverse-variance weights above, that is, only because the weights carry the $f_i^2$. The matrix $\mathbf{M}$ is therefore the inverse position covariance and no separate error propagation is required.

$\mathrm{Cov}(\mathbf{X})$ is expressed in ECEF, where ``horizontal'' and ``vertical'' are undefined, so we rotate it into the local East-North-Up frame at the object's own geodetic position. We then quote the $1\sigma$ semi-major axis of the East--North block as the horizontal error and the square root of the Up--Up element as the vertical error. We stress that this propagates \emph{random} pointing noise through the observing geometry alone: it carries no calibration bias, and is therefore a lower bound on the true position error rather than a prediction of the residual against ADS-B.

Once we compute time series of the track's 3D point, we consider different smoothing and fitting approaches that yield different velocity and acceleration estimations. The starting point is always the time series of 3D point in the ECEF reference frame, and in all cases, the velocity and accelerations are calculated as the numerical differentiation of the 3D point time series (once or twice, respectively). In the first approach, we consider a low degree polynomial curve fit of the 3D positions. Specifically, we consider linear and quadratic models, differentiated analytically once and twice; a linear fit therefore asserts a constant velocity and exactly zero acceleration, and a quadratic fit a constant acceleration. Second, we apply Savitzky–Golay smoothing (S-G, \citealt{10.1021/ac60214a047}) with degrees 1 and 2 to the 3D positions. We evaluate different smoothing windows: 21, 51, 101 and 201 samples. Finally, the last approach does not assume any global curve nor applies any smoothing to the positions; the velocities and accelerations are obtained by central differences taken on the recorded timestamps.

Because detections are missed intermittently, the sampling is not perfectly uniform, and we therefore evaluate the S-G polynomial in the samples' actual timestamps rather than through the equally spaced convolution coefficients. This reproduces the standard filter exactly when the frame interval is constant and remains correct when it is not, so no track has to be split at dropouts and no frames are discarded. 

\subsubsection{Track association}\label{sec:track_association}

\begin{figure}[t]
\centerline{\includegraphics[width=0.9\textwidth]{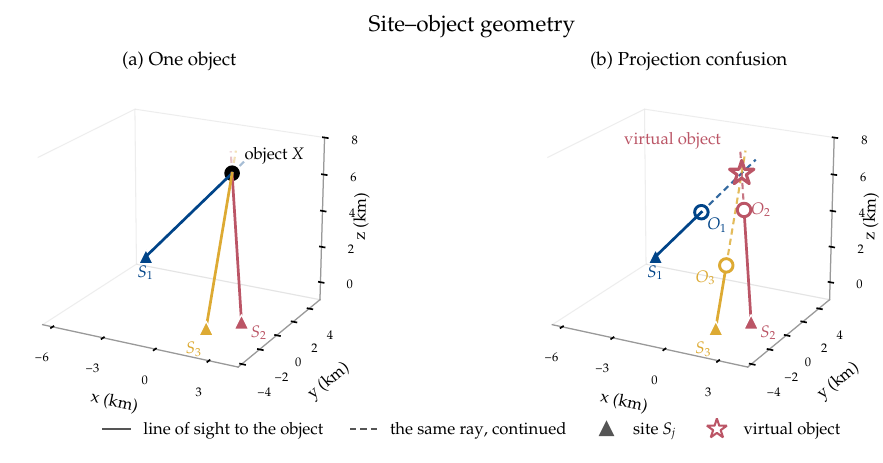}}
\caption{Track association by line of sight intersection diagram. Correctly associated detections of the same object \textbf{$X$} from different sites \textbf{$S_j$} are seen in the left panel. On the right panel, an unfavorable case where the intersecting lines of sight of different objects \textbf{$O_j$} locate a virtual object in 3D space.}
\label{fig:track_association_diagram}
\end{figure}

While triangulating objects' positions can be solved as a geometric problem, one of the greatest triangulation challenges is associating the same object as seen across different cameras. When we have objects with known ground truth positions (e.g. aircraft, the Moon or the Sun), we can first identify them in our observed tracks by projecting their 3D positions to the 2D camera pixel map and then match these to our detected tracks. Once identified, we can perform the triangulation of these objects if they share $id$s and overlap over time. However, this approach is restricted to identified objects, limiting the discovery of UAP. To extend the position triangulation to any object we present the strategy used to triangulate the position of any observed object.

To enable any object triangulation, we first need to associate the observed tracks that correspond to the same object. For this purpose, we perform a pairwise comparison of all tracks seen from cameras of different sites across an overlapping time window. To reduce the number of comparisons, we restrict to camera pairs that have a shared FoV.

Singling out objects seen simultaneously by different cameras poses the challenge of avoiding depth confusion, as objects with intersecting or nearly intersecting lines of sights may result in false positive matches. This projection confusion is inherent to the method and is amplified by the uncertainty of the sensor set up, since a near-coincidence falls inside the same tolerance that a genuine object does. As a result of mismatched detections, the position of a virtual object is calculated instead (see Fig.\,\ref{fig:track_association_diagram}). To reduce the number of spurious matches, we leverage the time-series information of the track to link tracks that have consistent trajectories over overlapping time windows.

For the comparison, we set a number of different gates that reduce the chance of matching two distinct objects. These include: temporal overlap, geometric agreement and velocity constraints. In detail, these gates are (in order):

\begin{enumerate}
    \item{\textbf{Temporal overlap}: The pair of tracks must overlap over a span of at least one second and share at least 75\% of the detections of the sparser track within that window; a pair failing this is still admitted if both tracks contribute at least 10 detections in the overlap window.}
    \item{\textbf{Coplanarity}: The two camera rays and the baseline vector must be \emph{nearly} coplanar. This condition ensures that the camera rays intersect, within the expected error given by the re-projection error of the camera combination. The threshold is set per camera pair as $\min\left(3\sqrt{\epsilon_{90^{th}\,A}^2+\epsilon_{90^{th}\,B}^2},\,0.05\right)$, with $\epsilon_{90^{th}}$ the $90$th-percentile extrinsic re-projection error expressed as an angle [rad]. The points where the camera rays are closest to each other should also be in front of the camera.}
    \item{\textbf{3D speed match}: If we follow the movement described by the points within the camera rays nearest each other, we can compute a 3D velocity of such points per camera. The velocity magnitudes must be within 20\% of each other, and their average velocity directions must have a cosine > 0.95. This aids the removal of spurious virtual objects.}
\end{enumerate}

These gates seem redundant for a real object, however they are aimed to restrict the wrong projection pairs: coplanarity is a single-frame test and by construction can easily be satisfied by a virtual object, whereas the speed and direction gates use the whole overlapping time window and are what actually break the degeneracy.

The pairs of tracks that pass these gates become matching candidates. As an object's track might be split in different segments in one of the cameras but not in the other, we allow multiple matches for a single track as long as their time windows do not overlap. The final associated tracks are selected to maximize the number of simultaneous detections and minimize the median re-projection error among all candidates. Pairwise matches are then closed transitively, so that a track matched to a second camera and independently to a third places all three in one object.

Multi-sensor fusion may also reduce spurious matches. While also subject to unfortunate projection confusion, a third (or more) sensor detecting the same virtual object is unlikely. Validating an object triangulation with three or more cameras depends, however, on advantageous observing geometry. In other words, that the object is observable from the three sites simultaneously. We consider pair-wise and three-site results in Sec.\,~\ref{sec:proper_triangulation}.

\section{Results}

In this section we present, on the one hand, the commissioning results of the Las Vegas IR Daleks and, on the other, the triangulation results. In the first, we focus on the results per camera, while in the second we focus on the analysis that requires an interconnection of the network.

\subsection{Commissioning results}\label{sec:commissioning}

\subsubsection{Detections and tracks}

\begin{figure}[t]
\centerline{\includegraphics[width=0.9\textwidth]{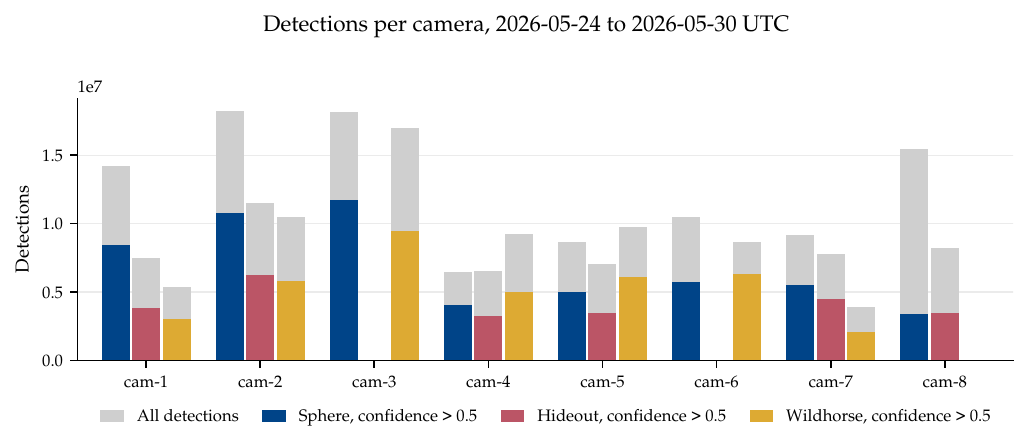}}
\caption{Detections per camera, 2026-05-24 to 2026-05-30 UTC. Grey is the total; the coloured portion is the subset above confidence $0.5$.}
\label{fig:detections_summary}
\end{figure}

The first part of our processing pipeline consists of running the YOLO detector and tracker on every recorded frame. Over the analyzed week the 21 active cameras produced $213{,}621{,}108$ detections, of which 55\% carry a confidence level above $0.5$. Grouping these in time yields $5{,}313{,}389$ tracks. Fig.~\ref{fig:detections_summary} shows the detection counts per camera and Fig.~\ref{fig:tracks_summary} the corresponding track counts.

\subsubsection{Object identification}\label{sec:identification}

\begin{figure}[t]
\centerline{\includegraphics[width=0.9\textwidth]{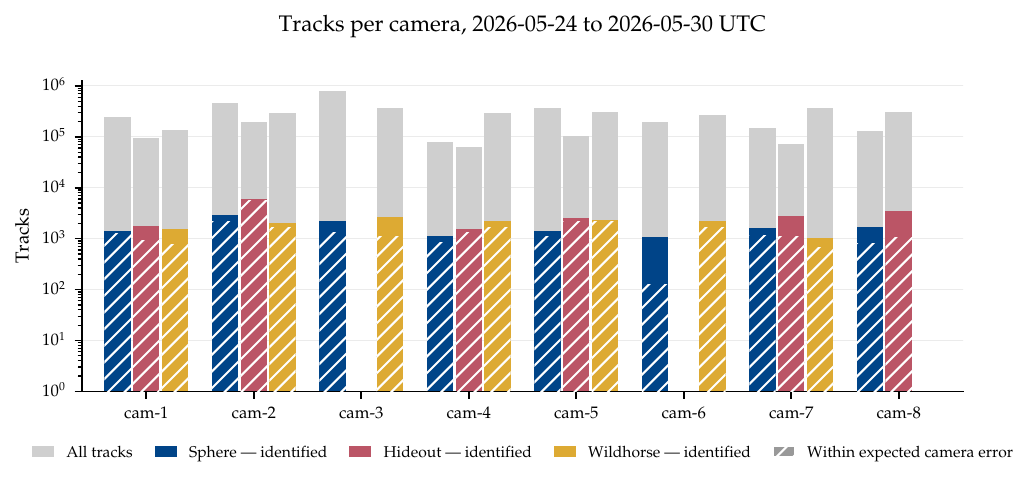}}
\caption{Tracks per camera over the analyzed week. In grey are the tracks that pass the confidence and length filter, while the coloured portion are those matched to an identified object. The white shades indicate the identified tracks found within the expected camera re-projection error.}
\label{fig:tracks_summary}
\end{figure}

While our primary goal is to study UAP, part of our analysis requires the identification of known objects for different purposes. On the one hand, it serves as a ground truth verification of the approach taken. As the nature of the objects is well-known, we can study the limits of our instruments. On the other hand, recognizing an identified object in our dataset discards the possibility of considering it a UAP.

Among the analyzed tracks, we found several natural objects, like birds, bees, leaves, the Sun or the Moon, and human-made objects like airplanes or helicopters. We do not report any sightings of satellites nor Solar System bodies, besides the Sun and the Moon. To avoid spurious identifications, we restrict the identifications to aircraft and solar system bodies. Additionally, we only test for position, velocities and acceleration using identified aircraft, as we can better test our method with a wider range of speeds, sizes, and distances.

From the $5,313,389$ relevant tracks analyzed, $44,417$ were matched to an identified object (Moon, Sun or aircraft): $41{,}337$ aircraft, $2{,}706$ Sun and $374$ Moon. This represents $0.84$\% of the total, and $70$\% of these were found within the expected camera re-projection error, meaning that the track's median residual falls within that camera's $90$th-percentile extrinsic re-projection error. Each camera detected over $1,000$ identifiable objects during this period, as seen in Fig.~\ref{fig:tracks_summary}.

\begin{figure}[t]
\centerline{\includegraphics[width=0.9\textwidth]{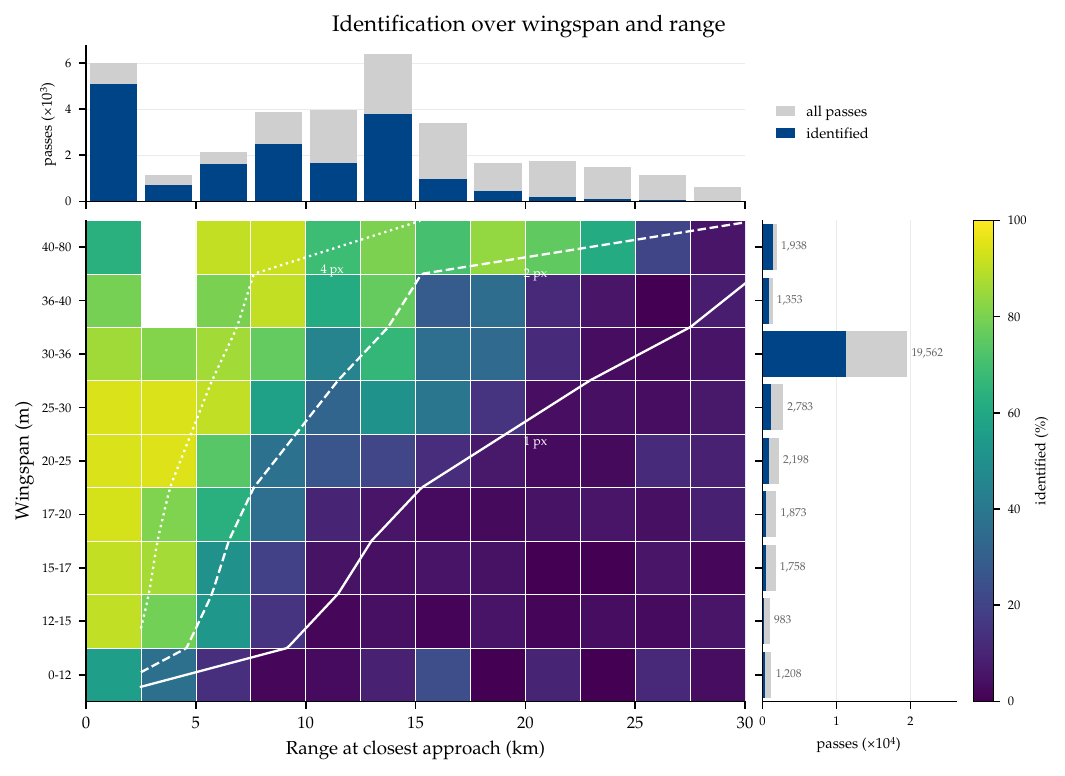}}
\caption{Percentage of identified aircraft per size and distance to detector. Cells with less than 10 elements are kept blank (total objects in blank cells: 13). Aircraft with apparent size smaller than $1\,pix$ is hardly identified by our cameras. The white contours show constant apparent sizes for the outward-looking cameras; the zenith cameras, whose pixels are $1.8\times$ coarser, need $1.8\times$ the wingspan to present the same apparent size.}
\label{fig:object_size_joint}
\end{figure}

We note that not all ADS-B transmitting aircraft that pass geometrically in the cameras' FOV within 30\,km during their recording time are detected. We analyze how the objects' detectability depends on their apparent size. For this, we calculate the percentage of aircraft passes that are identified per camera at the closest approach to the site and per aircraft's wingspan, as a size proxy\footnote{The aircraft wingspan is derived from ADS-B data, and could only be determined for 76\% of the aircraft geometrically in the field of view.}. As seen in Fig.\,\ref{fig:object_size_joint}, when the aircraft size in the camera is between 1-2 pixels, the detection is not certain. Above this limit, missed aircraft are rare, and under the 1 pixel limit any detection is also scarce: the identification fraction crosses $50$\% at an apparent wingspan of $1.8$ on-axis-equivalent pixels, falling to $1.5$ pixels once passes near the horizon are excluded. Roughly a fifth of the apparent size threshold is therefore atmosphere and terrain rather than target size, confirming the estimations from our design \cite{doi:10.1142/S2251171723400020}.

\subsection{Triangulation results}

All the triangulation results in this section correspond to tracks associated with the method presented in Sec.~\ref{sec:track_association}. In the case of identified objects, the identifications were transferred to the triangulated object if it was identified in at least two cameras contributing to the triangulation.

We use the mid-point position among all three sites to estimate relative distance uncertainties: $36.083680^{\circ}$ N, $115.102134^{\circ}$ W, 636.2 m.

\subsubsection{Triangulation coverage}\label{sec:coverage}

While each camera observes a solid angle of $40\,^{\circ}\,\times\,50\,^{\circ}$ (and $72\,^{\circ}\,\times\,95\,^{\circ}$ for zenith cameras), the combination of all the cameras from a Dalek array form a more complex volume (see Fig.\,\ref{fig:dalek_fov}). Using the intrinsic and extrinsic calibration of the IR-Daleks, we calculated the final sky coverage as the intersection of such volumes. In other words, the volume where triangulation could be geometrically performed if the object were observable from the respective sites.

\begin{figure}[t]
\centerline{\includegraphics[width=0.75\textwidth]{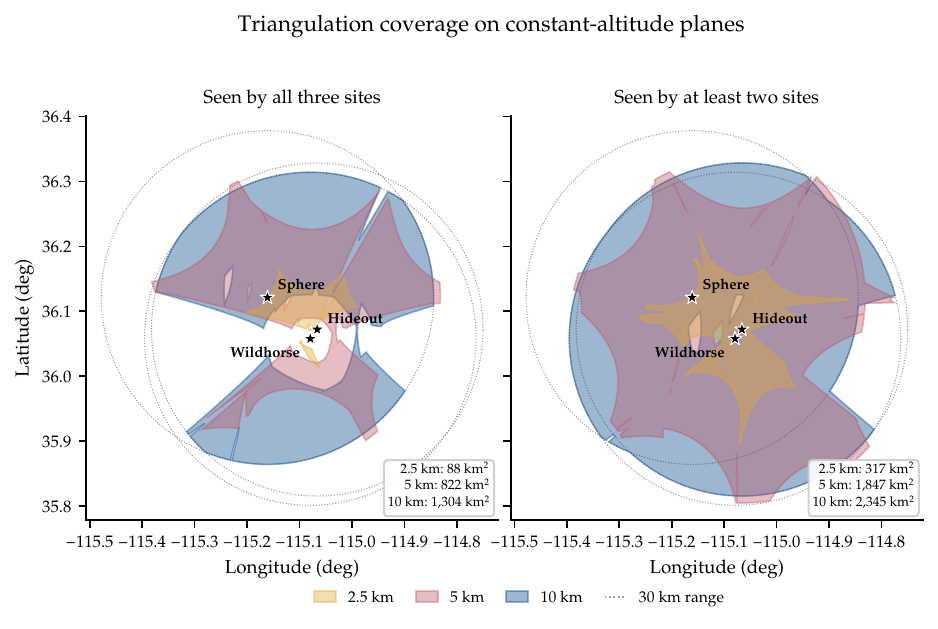}}
\caption{Combined geometric sky coverage of the Sphere, Hideout and Wildhorse sites at different altitudes. On the left-hand side panel, the intersection of the three sites' FoV. On the right-hand side, the FoV where there is coverage from at least two sites.}
\label{fig:spatial_coverage}
\end{figure}

To represent this volume, we used horizontal cuts at different altitudes: $2.5$\,km, $5$\,km and $10$\,km. For a 3-site triangulation, the volume is the intersection of each Dalek's FoV, while a pair-wise triangulation only requires an object to be seen by two cameras from different sites, resulting in a more extended common FoV as seen in Fig.\,\ref{fig:spatial_coverage}.

Due to the upward pointing of the cameras, the common FoV at lower altitudes is the most limited of the three heights shown. Conversely, at 10\,km the FoV coverage is only restricted by the 30\,km radius limit. It is worth noticing that for an object at greater distances, the camera resolution has a larger impact on the object detectability, and therefore on the ability of triangulating the position of such object.

\begin{figure}[t]
\centerline{\includegraphics[width=0.9\textwidth]{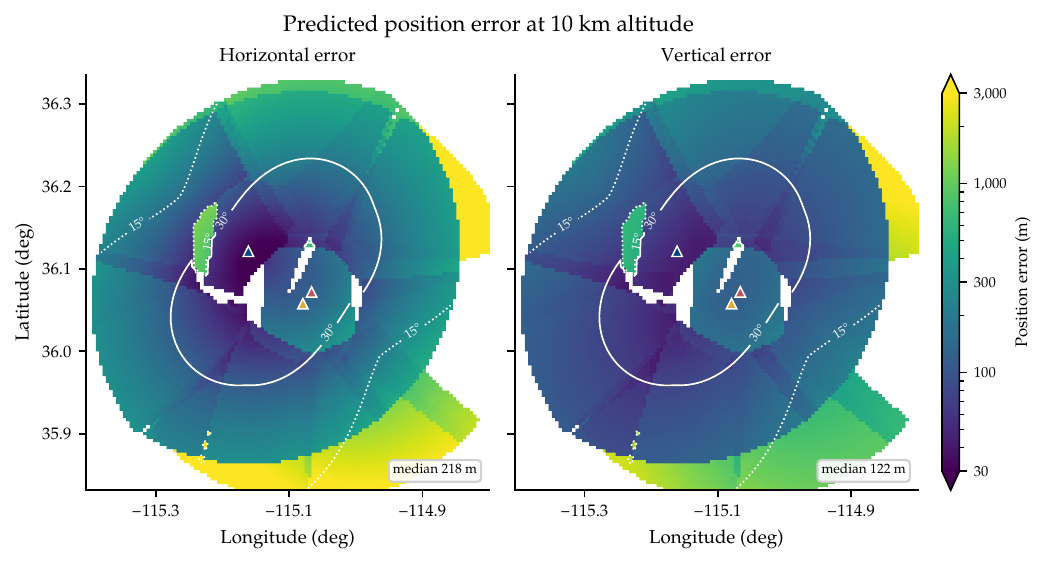}}
\caption{Predicted position error of a single triangulated detection on the $10\,$km
altitude plane, in metres: the horizontal error (left) and the vertical error (right), on a
common colour scale. Both are $1\sigma$, read off $\mathrm{Cov}(\mathbf{X})=\mathbf{M}^{-1}$
rotated into the local East--North--Up frame. The horizontal value is the semi-major axis
of the East--North block and the vertical value the square root of the Up--Up element. The
per-camera pointing noise driving them is that camera's median extrinsic re-projection
residual divided by its own focal length. White contours mark the best cross-site
ray-intersection angle and triangles the three sites; blank areas are seen by fewer than two
sites or lie beyond the $30\,$km detection range.}
\label{fig:position_error}
\end{figure}

Geometric coverage says where a position can be recovered at all; the same calibration also says how well. Evaluating $\mathrm{Cov}(\mathbf{X})=\mathbf{M}^{-1}$ of Sec.~\ref{sec:triangulation_math} on a grid over the $10\,$km plane gives the predicted $1\sigma$ position error of a single detection everywhere in the common field of view (Fig.~\ref{fig:position_error}). Over the region seen by all three sites the predicted horizontal error has a median of $230\,$m and a $90$th percentile of $0.6\,$km, whereas over the region seen by only two the median is $285\,$m and the $90$th percentile $4.3\,$km: the long tail of the error budget lives entirely in the two-site part of the volume, where the cross-site rays meet at a shallow angle. The vertical error is consistently the smaller of the two, at $0.56$ times the horizontal at the median, because the array's baselines are horizontal while the targets are seen well above the horizon. This is a statement about what the geometry and the calibration allow, not about what any particular detection achieved.

\subsubsection{Associated tracks}

\begin{table}[t]
  \centering
  \caption{Every stage from a pairwise triangulation to a distance that can be checked against ADS-B, over 2026-05-24 to 2026-05-30. The unit in the upper block is one \emph{pairwise triangulation} --- one object solved from one cross-site camera pair --- so an object seen by all three sites contributes three rows to it. The lower block counts objects instead. The median relative distance error over the scored sample is 0.014.}
  \label{tab:triangulation-funnel}
  \begin{tabular}{lrrr}
    \toprule
    Stage & $N$ & \% of pairwise & \% of parent \\
    \midrule
    Pairwise triangulations & 18,176 & 100.0 & -- \\
    \quad also solved from three sites & 2,213 & 12.2 & -- \\
    \quad identified in $\geq$1 camera & 9,894 & 54.4 & -- \\
    \qquad same object in both cameras & 5,773 & 31.8 & 58.3 \\
    \qquad named by one camera only & 3,917 & 21.6 & 39.6 \\
    \qquad conflicting names & 204 & 1.1 & 2.1 \\
    \quad aircraft scored against ADS-B & 5,627 & 31.0 & -- \\
    \qquad distance within 5\,\% & 4,681 & 25.8 & 83.2 \\
    \qquad distance 5--10\,\% & 500 & 2.8 & 8.9 \\
    \qquad distance beyond 10\,\% & 446 & 2.5 & 7.9 \\
    \midrule
    Distinct objects & 16,278 & -- & -- \\
    \quad solved from three sites & 739 & -- & -- \\
    \qquad with an ADS-B reference & 598 & -- & -- \\
    \qquad with $\geq$201 samples & 365 & -- & -- \\
    \bottomrule
  \end{tabular}
  \par\smallskip
  \parbox{0.92\linewidth}{\footnotesize\raggedright The last column is a percentage of the parent row. ``Also solved from three sites'' is a property of the object, not a step towards the accuracy rows: an aircraft is scored on its pairwise solution whether or not a third site saw it. The three name rows partition the identified count. Only a pair whose two cameras agreed carries a per-frame ADS-B position, so 5,605 of the scored matches come from the first of them and 22 from the third; the 3,917 named by a single camera receive no external check at all. Conflicting names are a lower bound on false association: they are only visible when both cameras named something.}
\end{table}

Applying the association scheme of Sec.~\ref{sec:track_association} to the analyzed week yields 18,176 cross-site matched track pairs, of which 2,213 are supported by all three sites and joined as 739 unique objects (Table~\ref{tab:triangulation-funnel}).

Of the total number of pairs, 9,894 (54.4\%) were identified in one or both cameras: in most cases (58.3\%) the tracks have the same id in both cameras, while 3,917 were only identified in one camera (39.6\%) and 204 have mismatching ids across cameras (2.1\%). These groups could be subject to a mislabeling of the object or a wrong track association. To solve this difference, we further compare the triangulated position to the matched ground truth aircraft in Sec.~\ref{sec:proper_triangulation}.

When only considering associated tracks from three sites, those identified as an airplane are 598 from the 739 total. This 19.1\% reduction contrasts with that of the pairwise track associations, where 45.6\% of the matches carry no identification at all.

\begin{figure}[t]
\centerline{\includegraphics[width=1\textwidth]{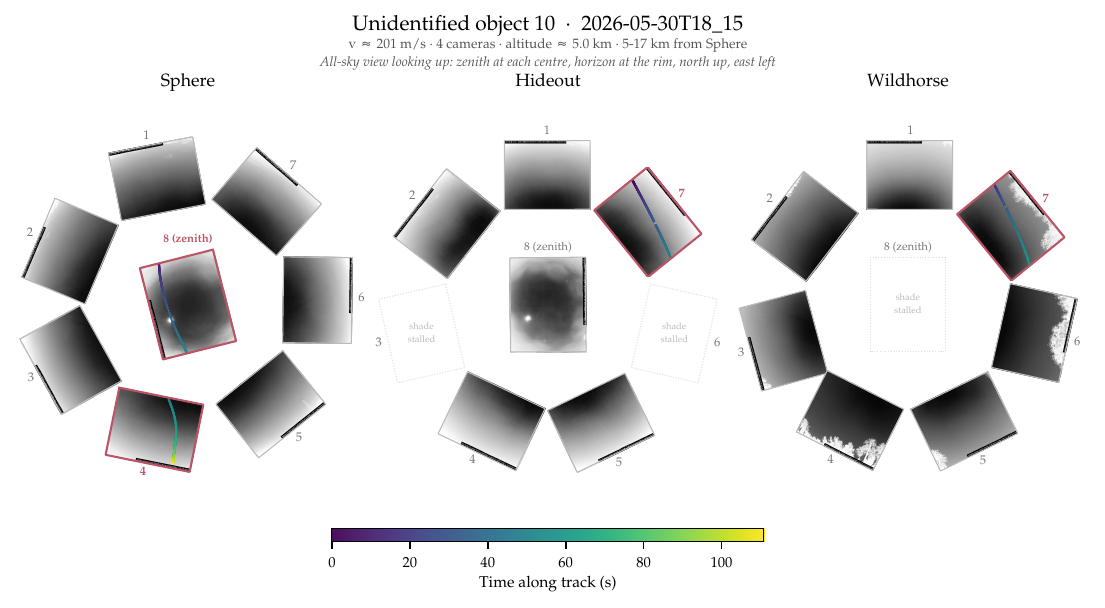}}
\caption{One object seen simultaneously across the network. Each panel is the view from one contributing camera at the same instant, with the detection visible in Sphere-cam-8 and cameras 7 at Hideout and Wildhorse.}
\label{fig:flying_object_example}
\end{figure}

The tracks are associated pairwise, however the further linkage of the tracks through transitive match allows us to recover tracks of a unique object across multiple cameras and sites.    Fig.~\ref{fig:flying_object_example} shows one object recovered across the three sites and seen in six cameras. It moves at $202$\,m/s and an average altitude of $5.0$\,km. With a median distance from the array midpoint of $6.9$\,km (closest approach $6.6$\,km), this object could not be identified with the available ground truth data. However, the trajectory shape, altitude and speed are compatible with an airplane.  

\subsubsection{Recovering triangulated positions, speeds and accelerations}\label{sec:proper_triangulation}

\begin{figure}[t]
\centerline{\includegraphics[width=0.9\textwidth]{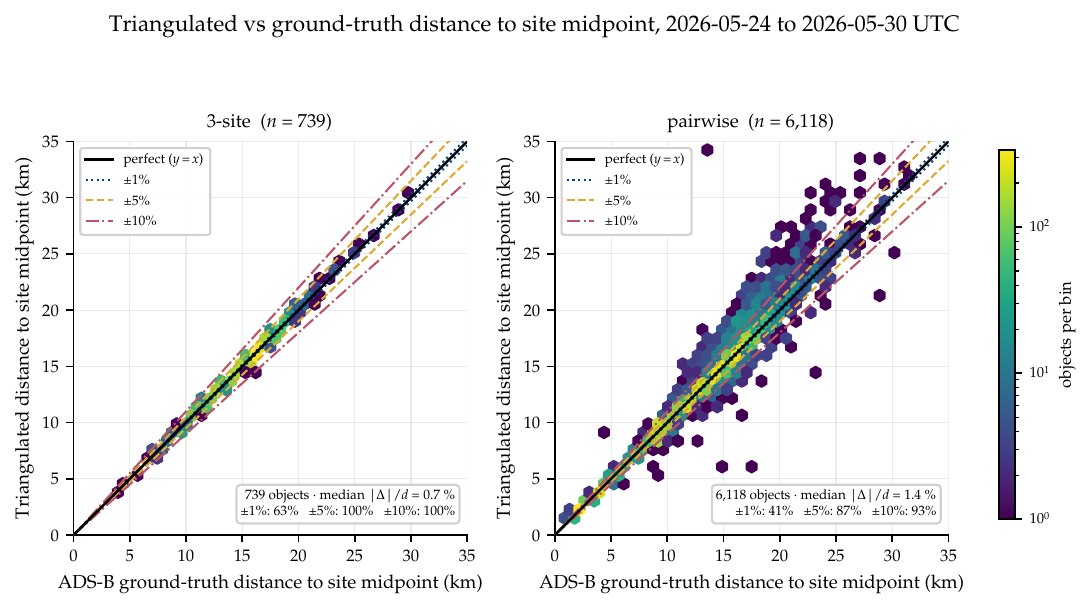}}
\caption{Triangulated distance against the distance implied by the ADS-B record, for identified aircraft. The diagonal is equality and the bands mark fixed fractional errors.}
\label{fig:recovering_aircraft_positions}
\end{figure}

\begin{figure}[t]
\centerline{\includegraphics[width=0.9\textwidth]{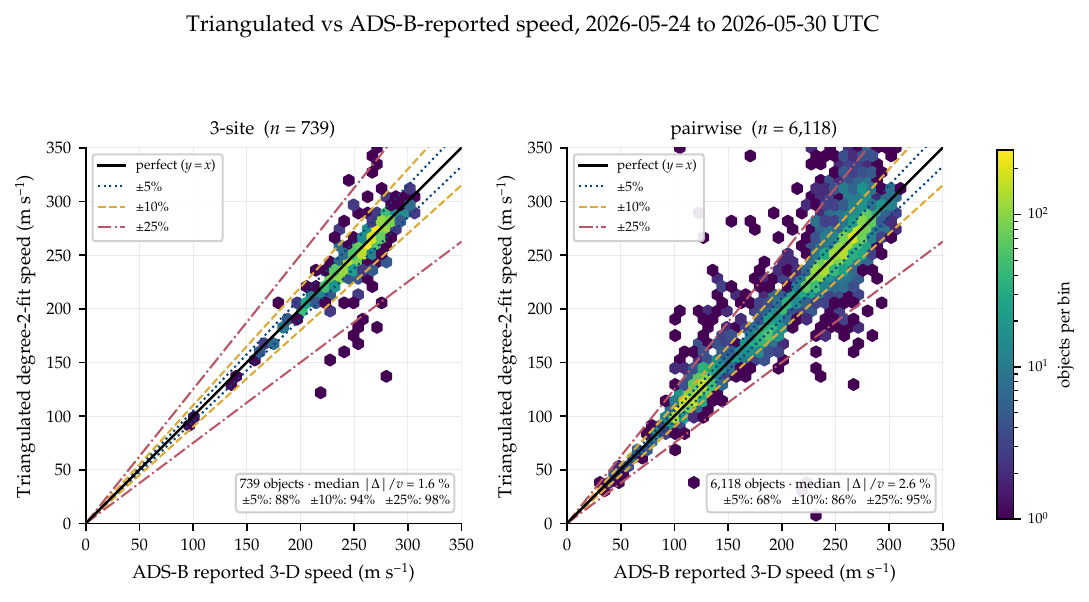}}
\caption{As Fig.\ref{fig:recovering_aircraft_positions}, the triangulated speed against the transponder's reported speed, for the same aircraft.}
\label{fig:recovering_aircraft_speeds}
\end{figure}

To measure the accuracy of the triangulated positions, velocities and speeds of objects, we compare to the ground truth provided by the ADS-B records of identified aircraft. 

In this section, we consider that an object has been correctly identified if it was identified simultaneously in at least two sites. This is aimed to avoid assuming a triangulation error, when this is the effect that we aim to analyze. Additionally, we require a minimum of 10 data points to calculate their dynamical properties.

Fig.~\ref{fig:recovering_aircraft_positions} compares the triangulated distance against the distance implied by the transponder, and Fig.~\ref{fig:recovering_aircraft_speeds} does the same for speed. In this case we consider the quadratic fit of the full 3D track to estimate the velocity. This allows us to extend the number of objects analyzed beyond those with $n-$samples, $n$ defined by the smoothing samples over the time window. In both figures, the left-hand-side panel shows triangulations where 3 sites contributed, while the right-hand-side one shows only pairs of sites triangulations. 

Among 5,773 pairs of tracks where the identification agrees across cameras, only 167 correspond to SSB. The ADS-B records further verify the airplane identifications and associated tracks as $83$\% of all pairs have been linked to an aircraft whose position has been retrieved within 5\% distance accuracy and 40\% show a distance accuracy within 1\%. The 3,917 pairs named by only one of the two cameras carry no per-frame ADS-B position at all, so they receive no external check: 5,605 of the 5,627 scored matches come from pairs whose two cameras agreed, and the remaining 22 from pairs with conflicting names.

Given that other objects like the Sun, Moon, or clouds can be correctly associated across cameras, but do not provide a reliable measure for verification, these values estimate an upper limit to how much projection contamination we can expect in the pairwise sample. 

99\% of all 3-site triangulated objects have been retrieved within 5\% of the ground truth distance, and 66\% of them within 1\%. 

When the triangulated positions are constrained by the three sites simultaneously, 588 of the 593 aircraft distances were estimated within $\pm\,5$\% of the ground truth ($99$\%), and 66\% of the objects are within $\pm\,1$\%. 

An analogue comparison regarding the objects' speeds recovered shows that most objects (78\%) get speeds recovered within $\pm\,5$\% of their ground truth value when the position calculation originates from the three sites, and only 5\% of the objects get speed errors beyond 10\% of the ground truth values. When only two sites contribute to the speed determination, {63}\% of the objects have speeds recovered within $\pm\,5$\%. We note that some objects have supersonic recovered speeds while this limit is never exceeded by the ground truth data: the fastest three-site solution reaches $637\,$m$\,$s$^{-1}$ and the fastest pairwise one is unphysical, against a maximum reported ADS-B speed of $308\,$m$\,$s$^{-1}$. 

The pairwise overestimates are not spread evenly over the array. Of the $5{,}626$ scored pairwise speeds, $614$ ($11$\%) exceed the transponder value by more than $25$\%, and $93$\% of those come from the Hideout--Wildhorse pair, which supplies only $45$\% of the sample. Their $2\,$km separation is five times shorter than the other two baselines and admits proportionally more range error; the distance residuals of that pair are correspondingly wider but stay symmetric, while speed, as the magnitude of a velocity vector, rectifies that scatter upward and so places it above the diagonal.

\begin{figure}[t]
\centerline{\includegraphics[width=1\textwidth]{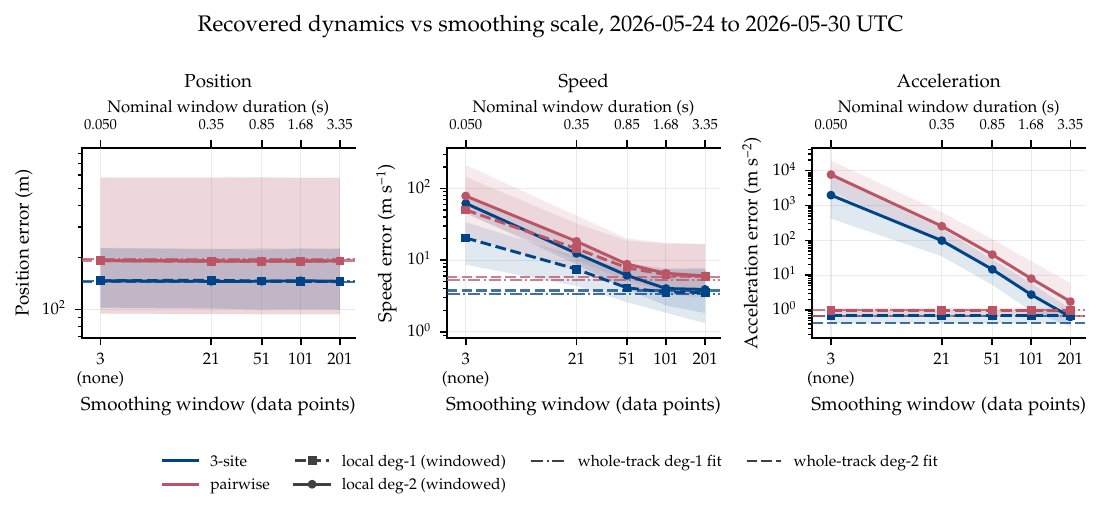}}
\caption{Median error in recovered speed and acceleration against smoothing scale, for the local polynomial estimators of degree 1 and 2, the whole-track linear and quadratic fits, and unsmoothed differentiation. All levels are measured on the same objects so that the comparison reflects the estimator and not a change of population. Position and speed: 3-site $n=292$, pairwise $n=5{,}530$; acceleration, scored only where the transponder reported a velocity change: 3-site $n=262$, pairwise $n=5{,}171$.}
\label{fig:smoothing_tracks}
\end{figure}

A quadratic fit is normally a sensible choice for an aircraft track being followed for up to 1 minute. However, as we aim to extend the analysis we further explored how accurately the kinematic properties of such aircraft can be retrieved when smoothing the raw tracks with a linear or quadratic local fit. Furthermore, the results of different smoothing windows are compared in Fig.~\ref{fig:smoothing_tracks}.  This figure shows how the recovered kinematics depend on the estimator and on the smoothing scale, for the whole family of estimators described in Sec.~\ref{sec:triangulation_math}.

As the smoothing is performed only on the positions, the averaged positions do not reflect any improvement from the smoothing. However, the speed and acceleration average residuals are visibly reduced as the smoothing window increases. For reference, we include the residual between the global fits and the ground truth tracks, which sets a limit to the track recovery based on our system set up. The locally-linear curves are flat in the acceleration panel because a degree-1 fit asserts $a\equiv0$ by construction, so what they trace is the magnitude of the reference acceleration itself ($0.71\,$m$\,$s$^{-2}$ for the three-site sample, $0.99\,$m$\,$s$^{-2}$ pairwise) and not an estimator result.

The kinematic residuals, when retrieved by differences from the raw data, are large: a median speed error of $67\,$m$\,$s$^{-1}$ for the three-site sample and $80\,$m$\,$s$^{-1}$ pairwise, against targets moving at $94$--$307\,$m$\,$s$^{-1}$, and acceleration fluctuations of $\sim5{,}100$ and $\sim5{,}600\,$m$\,$s$^{-2}$ respectively, against recovered values that should be close to $0$--$1\,$m$\,$s$^{-2}$. The speed retrieval stabilizes in the 101-201 window samples regime under 10\,m/s ($7.1\,$m$\,$s$^{-1}$ three-site and $9.0\,$m$\,$s$^{-1}$ pairwise at $w=201$, degree 2), for aircraft that usually move at hundreds of meters per second. The accelerations, although greatly reduced after smoothing, still sit above the reference acceleration at the 201-sample window, at $0.69\,$m$\,$s$^{-2}$ three-site and $1.88\,$m$\,$s$^{-2}$ pairwise against $0.51$ and $0.80\,$m$\,$s$^{-2}$ for the whole-track quadratic; the three-site curve has, however, essentially converged by that window.

\subsection{Triangulated objects during 2026-05-24 to 2026-05-30 UTC}

\begin{figure}[t]
\centerline{\includegraphics[width=1\textwidth]{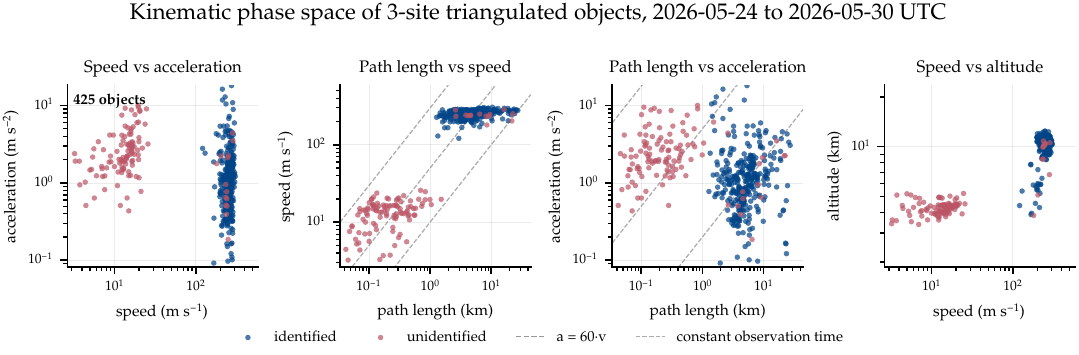}}
\caption{Kinematic phase space of the $365$ objects solved from all three sites with at
least $201$ samples, in four projections of the same table: speed against acceleration, path
length against speed, path length against acceleration, and speed against altitude. Speed
and acceleration come from a local degree-2 polynomial over $201$ samples fitted to the raw
triangulated positions. Objects linked to a ground-truth source are drawn apart from those
that were not. Dashed lines mark $a=60\,v$, the ceiling of what differencing a $60\,$fps
track can produce, and the constant-observation-time diagonals.}
\label{fig:phase_space}
\end{figure}

\begin{figure}[t]
\centerline{\includegraphics[width=1\textwidth]{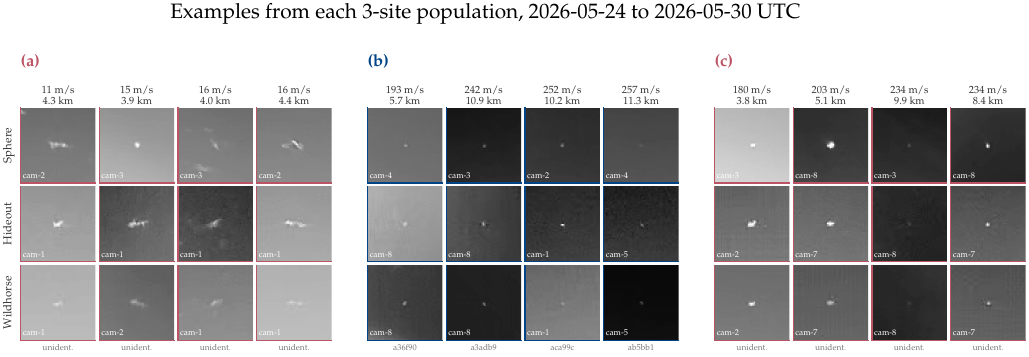}}
\caption{Cutouts of examples of triangulated objects, as seen from the three sites (from top to bottom, Sphere, Hideout and Wildhorse). The panels show, from left to right, unidentified objects with low speeds (<30\,m/s), identified aircraft and unidentified objects with high speeds (>100\,m/s). For each object, the speed and distance to the sites' midpoint are indicated above the object's cutouts.}
\label{fig:triangulated_examples}
\end{figure}

We consider now all the objects for which we could reliably calculate their positions and speeds, without assuming a global trajectory. These are 365 tracks that have been simultaneously observed from the three sites and each 3D track has over 201 points.

With the goal of finding true outliers, we relaxed the object identification requirements from those used in Sec.~\ref{sec:proper_triangulation}. Here, we consider that the object is correctly identified if at least one camera track was linked to a ground truth object, where the triangulation error is within $5\,$km.

To characterize the triangulated tracks, we display the average kinematic properties of such objects in four different planes: \emph{i)} speed vs acceleration, \emph{ii)} path length vs speed, \emph{iii)} path length vs acceleration, and \emph{iv)} speed vs altitude (Fig.~\ref{fig:phase_space}). The tracks span in speeds up to 298\,m/s at the 99th percentile. Only two objects exceed 306\,m/s: one reaching 993\,m/s carries a single-camera label whose aircraft lies 9\,km from the triangulated track, and is discarded as a mismatch, while the other, at 377\,m/s, is named in agreement by all three sites, sits 435\,m from that aircraft and still overshoots its transponder speed of 267\,m/s, for the reason given in Sec.~\ref{sec:discussion}. Neither is a fast target. Furthermore, although mostly under 5\,m/s$^2$ (25 of the 365 exceed it), accelerations reach 17.5\,m/s$^2$ at the 99th percentile. The tracks were followed up to $28\,$km (median $4.1\,$km) and were seen at altitudes between 3.1 and 12.5\,km.

We found that two distinct populations separate from each other in the kinematic phase space. One of the groups was primarily identified aircraft (284), although we also found 17 unidentified objects that have similar features. The other group had only objects that could not be linked to known aircraft or SSBs (64). After inspection, we found that these objects were small clouds (see examples in Fig.~\ref{fig:triangulated_examples}). The separation is clean rather than merely visual: no object at all in this sample has a median speed between $30$ and $100\,$m$\,$s$^{-1}$.

The identified aircraft show speeds between 116\,m/s and 298\,m/s, while the clouds move in agreement with wind speeds and remain under 30\,m/s. This speed difference is also reflected in the path length difference, where the aircraft are tracked for over a kilometer (median $4.8\,$km) while the clouds are rarely followed beyond 1\,km (median $0.18\,$km). The small clouds are normally moving at 3 to 5\,km of altitude, while aircraft are mostly seen at the cruise altitude of around 10\,km(median $10.5\,$km).

\section{Discussion}\label{sec:discussion}

The previous analyses allowed us to illustrate and quantify the capabilities of our triangulation sites. We highlight that geometric coverage does not guarantee visibility, as other factors may be involved; the object may not be resolved, there might be other objects obstructing the view (e.g. trees) or unfavorable weather conditions like clouds or haze.

We note that the sites had reduced cloud coverage over the week of data analyzed. However, weather variations may affect the object detection and triangulation completeness. Other seasonal variations, as the varying angle of the Sun combined with the protecting shades scheduling, could introduce additional yearly fluctuations. Furthermore, the instruments may experience position shifts over time due to wind or precipitations. While more data is necessary to report the sites' changes over time, this work provides a snapshot characterization of the sites and their capabilities for triangulation.

While aircraft detections under 1 pixel are possibly linked to misidentifications with another object in the line of sight, larger than 2 pixels aircraft might be missed by obstructions in the line of sight, like clouds or trees. The aircraft identification completeness can also vary over time due to the traffic mix.

In the search for UAP, we prioritize not assuming any global trajectory when triangulating. While this avoids imposing a given trajectory for an object, it comes at the cost of retrieving noisy tracks. In consequence, a traditional differentiation calculation of velocity and acceleration results in an overestimation of such values. We have shown that applying a simple smoothing function results in a better recovery of aircraft velocities and accelerations. When the smoothing window is large enough, we can recover kinematics comparable to those that assume a linear or quadratic curve for aircraft trajectories. This way, we identify what errors we can expect from the setup and calibration versus velocity differentiation.

Our triangulation approach differs from that proposed in \cite{doi:10.1142/S2251171723400020}, in that it weights each line of sight by its own pointing uncertainty and range rather than treating all cameras alike. Furthermore, in this work we implement a track smoothing strategy to reduce the speed and acceleration overestimations due to noisy tracks. A wide window carries one caveat of its own. When the set of cameras solving a track changes mid-pass, the solution steps between combinations whose systematic offsets differ, and a window spanning the change absorbs that step as motion. The fastest identified object of Fig.~\ref{fig:phase_space} is solved successively by three camera combinations offset by $290$, $458$ and $521\,$m from the aircraft's ADS-B position: a degree-2 fit over $201$ samples returns $377\,$m$\,$s$^{-1}$ against a reported $267\,$m$\,$s$^{-1}$, while the same fit applied to the ADS-B positions returns $274\,$m$\,$s$^{-1}$ and removing each combination's mean offset returns $271\,$m$\,$s$^{-1}$. Narrower windows sit inside a single combination and are unaffected ($271$--$280\,$m$\,$s$^{-1}$ at $w=51$--$101$). The effect is worst on fragmented tracks, where $201$ samples span $11.6\,$s of real time rather than the $3.3\,$s a gap-free track would give, so the window reaches across several combinations. While this results in a reliable estimation of speeds, accelerations may still deviate from the real values by $\sim0.7$\,m/s$^2$ for three-site solutions and $\sim1.9$\,m/s$^2$ for pairwise ones at the widest window we tested. While most accelerations were well under this value, some reconstructed trajectories may be affected by apparent jumps where an aircraft crossed from one camera's field of view into the next. Moreover, cloud tracks may introduce further variability to their reconstructed tracks, as the detection center could quickly change due to their shape shifting over time.

The data is currently processed in distinct stages, after the full recording is complete. A future extension could instead perform triangulation in real time. The S-G smoothing step is unsuited to this setting, since its window is centered on the sample being estimated and therefore requires future detections that are not yet available online. Recursive Bayesian filters (\citealt{Sarkka_2013}, \citealt{5589113}) offer a natural alternative: rather than fitting a polynomial to a completed position series, the Kalman filter and its variants perform the same estimation recursively, updating position, velocity, and acceleration as each new detection arrives, in effect acting as an online form of the same regression. Adopting such a filter would also make the track association step correspondingly more time-critical, since candidate matches would need to be scored and resolved as new track points are detected, rather than in a single batch pass over the full observation window — motivating the shift toward a probabilistic association scheme, that can assign or update a match incrementally as new points arrive.

The current association scheme uses a sequence of hard thresholds to discard candidates. A future enhancement of the pipeline may replace this stage with a probabilistic data association (\citealt{1102177}, \citealt{bar2011tracking}), replacing the binary decision with a continuous probability.

At the Galileo site in Las Vegas, there are additional wide field visible cameras that could complement the triangulation across these sites, both in wavelength range as well as extending the multi-sensor approach \cite{multi-fusion} to validate and better constrain the positions of the triangulated objects \cite{Hall2002}. Also arranged in dome-like structures, the visible mini Daleks have four cameras covering a $\sim85\,^{\circ}\times67\,^{\circ}$ field of view each, on $4128\times3008\,$px sensors pointing north, east, south and west. With the upcoming commissioning of the visible mini-Daleks, we expect to produce detections and tracks files that are directly comparable with those of the IR-Daleks. From this common ground, we can directly add these sensors' detections to the triangulation pipeline.

\section{Conclusions}

We have commissioned three long-wave infrared camera arrays near Las Vegas and built a triangulation pipeline on them that makes no assumption about the shape of an object's trajectory. Our main conclusions are:

\begin{enumerate}
    \item The array recovers three-dimensional positions of objects seen simultaneously from two or more sites by minimizing the weighted squared distance to the lines of sight at each instant, weighting each camera by its own pointing uncertainty and by its range to the target. Validated against ADS-B aircraft, this recovers distances within $5$\% for $99$\% of the $593$ aircraft solved from three sites ($66$\% of them within $1$\%) and for $83$\% of the $5{,}626$ solved from a single site pair ($40$\% within $1$\%), at a median relative distance error of $0.014$; speeds are recovered within $5$\% for $78$\% and $63$\% of them respectively.
    \item Not assuming a global trajectory model comes at the cost of noisier per-point estimates, and naive differentiation of those points overestimates speed and acceleration (by $67\,$m$\,$s$^{-1}$ and $\sim5{,}100\,$m$\,$s$^{-2}$ at the median for three-site tracks). A local polynomial smoothing of sufficient window recovers kinematics comparable to those obtained by imposing a linear or quadratic trajectory (on objects where this model is suitable), without imposing one: a degree-2 fit over $201$ samples reaches $7.1\,$m$\,$s$^{-1}$ and $0.69\,$m$\,$s$^{-2}$, against $6.0\,$m$\,$s$^{-1}$ and $0.51\,$m$\,$s$^{-2}$ for a whole-track quadratic.
    \item The array's completeness is set by apparent size, with the $50$\% identification threshold at $1.8$ on-axis-equivalent pixels of wingspan, but is strongly modulated by elevation and by how long a target dwells in the frame. Geometric coverage is therefore an outside bound on what can be identified, not a prediction of it.
    \item The triangulated population separates in kinematic phase space without any appeal to identification. Of the $365$ objects solved from three sites with at least $201$ samples, $301$ sit in an aircraft-like locus at $116$--$298\,$m$\,$s$^{-1}$ and a median altitude of $10.5\,$km  ($284$ of them confirmed by ADS-B and $17$ unidentified) while $64$ unidentified objects sit under $30\,$m$\,$s$^{-1}$ at $3$--$5\,$km and are, on inspection, small clouds. No object falls between the two groups, so the array separates the dominant populations on dynamics alone.
    \item Any statement about the population of unidentified objects must be made relative to a stated size, range and contrast, since what the array can detect at all is set by apparent size and by thermal contrast against the sky. 
    \item No object appears anomalous with a hypersonic speed or high acceleration during the analyzed week. All objects reliably tracked appear ordinary natural or human-made ones. The triangulation results are nonetheless promising: the accuracy demonstrated here is what would be required to support a defensible claim about an anomalous trajectory, were one to be found. Analysis of the recorded archive is ongoing.
\end{enumerate}

The natural extension of this work is to fold the visible-wavelength mini-Daleks into the same association and triangulation pipeline, which both extends the wavelength coverage and adds independent lines of sight to break the projection degeneracies described in Sec.~\ref{sec:track_association}.

\vspace{6pt}

\authorcontributions{Conceptualization, A.L., R.S. and A.D.; methodology, A.L. and R.S.; software, A.L., R.S. and E.M.; formal analysis, A.L. and R.S.; resources, A.D. and E.K.; data curation, A.L. and R.S.; writing---original draft preparation, A.L., R.S. and A.D.; writing---review and editing, A.L., R.S., A.D. and A.L.; supervision, A.L.; project administration, E.K. and A.L.; funding acquisition, A.L. All authors have read and agreed to the published version of the manuscript.}

\funding{This research was funded by the Galileo Project which benefits from generous donations by MSG Sphere, Eugene Jhong, Dr. Frank Laukien, the Richard King Mellon Foundation and the John Templeton Foundation.}

\institutionalreview{Not applicable.}

\dataavailability{The raw data supporting the conclusions of this article will be made available by the authors on request.}

\acknowledgments{We thank Dr. Matthew Szenher for his invaluable help with the software and guidance.}

\conflictsofinterest{The authors declare no conflicts of interest.}

\newpage
\abbreviations{Abbreviations}{
The following abbreviations are used in this manuscript:
\\

\noindent
\begin{tabular}{@{}ll}
ADS-B & Automatic Dependent Surveillance--Broadcast\\
ECEF & Earth-Centered, Earth-Fixed\\
FoV & Field of view\\
HFoV & Horizontal field of view\\
LWIR & Long-wave infrared\\
NTP & Network Time Protocol\\
PPS & Pulse per second\\
SDR & Software-defined radio\\
S-G & Savitzky--Golay\\
UAP & Unidentified aerial phenomena\\
UTC & Coordinated Universal Time\\
\end{tabular}
}

\appendixtitles{} \appendixstart
\appendix
\section[\appendixname~\thesection]{: IR-Dalek extrinsic calibration}\label{app:extrinsic-calobration}

Following the iterative strategy for the IR-Dalek extrinsic calibration (see Sec.~\ref{sec:dalek_calibration}), we present further details on the metrics used, choices made and results. In total, the calibration required three iterations to find minimal improvement between the last two results. 

Each iteration outputs a \emph{structural} and a \emph{direct} set of calibrations. The structural calibration is used as a base for the next calibration. The \emph{direct} calibration is the individual camera solve, while \emph{structural} adds the rigid-body fallback for cameras above 5\,px. Whenever a camera re-projection regresses, the base calibration is used instead.

As different extrinsic calibrations may result in different aircraft identified, we used a fixed sample of identified tracks to evaluate the different extrinsic calibrations. This sample is the union of all the identified aircraft with the different calibrations, which were verified by being simultaneously detected by two or more cameras and triangulated within 1km of the ADS-B position.

Table~\ref{tab:calibration-comparison} shows the median and 90$^{th}$ percentile of both re-projection and triangulation errors. The \emph{direct} calibrations are more accurate than the structural, however the structural keep improving over the iterations. The percentage of identified aircraft versus the fixed sample also increases over iterations, until a $\sim78\%$ plateau.

The best results lie between the second and third \emph{direct} iteration, with better tail behavior in the second iteration. Most cameras have almost identical rotation matrices between these configurations, however cam-5 at Hideout deviates to a worse calibration in the third iteration. For these reasons, we use the \emph{direct} second iteration for the rest of the analysis, whose details per camera can be found in Table~\ref{tab:it2-direct-per-camera}.

\begin{table}[H]
  \centering
  \caption{Extrinsic iterative calibrations fitted on 2026-05-22 and 2026-05-23, scored on one fixed sample of verified aircraft tracks. Best value per column is shown in bold.}
  \label{tab:calibration-comparison}
  \begin{tabular}{lrrrrr}
    \toprule
    && \multicolumn{2}{c}{Reprojection (px)} & \multicolumn{2}{c}{Triangulation (m)} \\
    \cmidrule(lr){3-4} \cmidrule(lr){5-6}
    Configuration & Ident. (\%) & Median & 90th & Median & 90th \\
    \midrule
    Design & 51.7 & 27.10 & 44.57 & 624 & 1,947 \\
    \addlinespace
    Iteration 1, direct & 76.2 & 4.33 & 20.94 & 212 & 1,164 \\
    Iteration 1, structural & 68.1 & 5.69 & 48.08 & 303 & 2,514 \\
    \addlinespace
    Iteration 2, direct & 77.9 & 4.07 & \textbf{13.66} & \textbf{177} & \textbf{1,082} \\
    Iteration 2, structural & 73.4 & 4.98 & 31.61 & 225 & 2,206 \\
    \addlinespace
    Iteration 3, direct & \textbf{78.8} & \textbf{4.05} & 18.80 & 179 & 1,850 \\
    Iteration 3, structural & 76.4 & 4.17 & 30.55 & 211 & 2,507 \\
    \bottomrule
  \end{tabular}
  \par\smallskip
\end{table}

\begin{table}[H]
\begin{adjustwidth}{-\extralength}{0cm}
  \centering
  \caption{Per-camera performance of the iteration-2 direct calibration, scored on the same verified sample as Table~\ref{tab:calibration-comparison}. $\Delta\theta$ is the angle between each camera's calibrated optical axis and its design orientation.}
  \label{tab:it2-direct-per-camera}
  \small
  \begin{tabular}{llrrrrrrrrrr}
    \toprule
    &&&&&& \multicolumn{2}{c}{Reproj. (px)} & \multicolumn{2}{c}{Triang. (m)} && \\
    \cmidrule(lr){7-8} \cmidrule(lr){9-10}
    Site & Cam. & Pairs & Ident. (\%) & Aircraft & ADS-B pts. & Med. & 90th & Med. & 90th & Elev. ($^\circ$) & $\Delta\theta$ ($^\circ$) \\
    \midrule
    Sphere & 1 & 83 & 73.5 & 53 & 2,086 & 5.06 & 12.46 & 535 & 800 & 27.3 & 2.76 \\
     & 2 & 403 & 92.8 & 229 & 8,736 & 2.81 & 6.59 & 169 & 338 & 28.3 & 1.81 \\
     & 3 & 174 & 92.0 & 108 & 3,861 & 4.88 & 7.88 & 158 & 392 & 30.8 & 2.33 \\
     & 4 & 163 & 85.9 & 76 & 3,111 & 4.56 & 7.99 & 145 & 757 & 34.3 & 5.48 \\
     & 5 & 197 & 93.4 & 119 & 3,925 & 3.45 & 7.08 & 62 & 150 & 30.1 & 0.85 \\
     & 6 & 5 & 100.0 & 5 & 171 & 1.90 & 3.96 & 154 & 469 & 32.7 & 2.67 \\
     & 7 & 11 & 72.7 & 7 & 347 & 7.25 & 19.68 & 197 & 1,372 & 29.4 & 0.92 \\
     & 8 & 82 & 95.1 & 59 & 2,299 & 2.79 & 6.25 & 128 & 231 & 86.4 & 3.47 \\
    \midrule
    Hideout & 1 & 481 & 84.8 & 190 & 4,164 & 12.65 & 19.93 & 193 & 629 & 31.8 & 1.99 \\
     & 2 & 416 & 69.5 & 148 & 2,457 & 9.11 & 16.42 & 270 & 2,037 & 29.0 & 0.92 \\
     & 3 & \multicolumn{10}{c}{excluded (shade stalled closed)} \\
     & 4 & 378 & 90.7 & 212 & 6,104 & 3.88 & 8.30 & 163 & 539 & 30.8 & 0.63 \\
     & 5 & 481 & 28.9 & 111 & 1,591 & 20.44 & 63.15 & 889 & 1,981 & 27.6 & 2.90 \\
     & 6 & \multicolumn{10}{c}{excluded (shade stalled closed)} \\
     & 7 & 225 & 84.9 & 93 & 1,474 & 5.57 & 9.66 & 106 & 249 & 30.7 & 1.20 \\
     & 8 & 184 & 86.4 & 96 & 1,829 & 4.70 & 8.25 & 147 & 233 & 89.4 & 0.79 \\
    \midrule
    Wildhorse & 1 & 178 & 83.7 & 81 & 1,385 & 9.42 & 12.91 & 114 & 336 & 32.6 & 2.86 \\
     & 2 & 347 & 76.9 & 166 & 4,550 & 4.03 & 9.55 & 162 & 537 & 31.2 & 1.75 \\
     & 3 & 151 & 90.1 & 96 & 2,635 & 6.01 & 9.85 & 120 & 440 & 31.2 & 3.08 \\
     & 4 & 394 & 90.6 & 210 & 4,206 & 4.82 & 12.26 & 234 & 700 & 26.9 & 3.38 \\
     & 5 & 370 & 50.5 & 135 & 4,358 & 1.84 & 4.11 & 798 & 1,951 & 32.5 & 2.38 \\
     & 6 & 265 & 86.4 & 127 & 3,900 & 2.03 & 5.27 & 68 & 180 & 32.0 & 1.93 \\
     & 7 & 183 & 88.5 & 58 & 707 & 2.04 & 6.13 & 105 & 351 & 31.0 & 1.45 \\
     & 8 & \multicolumn{10}{c}{excluded (shade stalled closed)} \\
    \midrule
    All &  & 5,171 & 77.9 & 579 & 63,896 & 4.07 & 13.66 & 177 & 1,082 &  & 1.99 \\
    \bottomrule
  \end{tabular}
  \par\smallskip
  \parbox{0.92\linewidth}{\footnotesize\raggedright \emph{Pairs} is the number of verified observed-track to ADS-B pairs for the camera; \emph{Ident.} the share of them this calibration itself identified and verified, and \emph{Aircraft} and \emph{ADS-B pts.} count the distinct aircraft and the ADS-B reports within those tracks. Reprojection and triangulation errors are computed over all the camera's verified pairs. A triangulated instant is attributed to every camera that contributed a ray, so the \emph{All} row counts each instant once and is not the mean of the camera rows; its $\Delta\theta$ is the median over cameras. \emph{Elev.} is the calibrated optical axis above the local horizon.}
\end{adjustwidth}
\end{table}

\begin{adjustwidth}{-\extralength}{0cm}

\reftitle{References}

\PublishersNote{}
\end{adjustwidth}
\end{document}